\documentclass[11pt,a4paper]{article}

\usepackage[colorlinks=true, linkcolor=black!50!blue, urlcolor=blue, citecolor=blue, anchorcolor=blue]{hyperref}
\usepackage[font=small,labelfont=bf,margin=0mm,labelsep=period,tableposition=top]{caption}
\usepackage[a4paper,top=1.5cm,bottom=2cm,left=2.5cm,right=2.5cm,bindingoffset=0mm]{geometry}

\usepackage[utf8]{inputenc}
\usepackage{graphicx}
\usepackage{epsfig}
\usepackage{cite}
\usepackage{multirow}
\usepackage{amssymb}
\usepackage{amsmath}
\usepackage{dsfont}
\usepackage{url}
\usepackage{xcolor}
\usepackage{afterpage}
\usepackage{url}
\usepackage{hyperref}
\usepackage{booktabs}
\usepackage{tabularx}
\usepackage{enumitem}
\usepackage{cite}
\usepackage{bm}

\newcommand{\myparagraph}[1]{\vspace{0.2cm} \noindent \textbf{#1} --} 

\begin{document}

\begin{flushright}
Nikhef-2020-041
\end{flushright}

$\qquad$
\vspace{0.2cm}
\begin{center}
  {\Large \bf Self-consistent determination of proton and nuclear PDFs\\[0.3cm] at the Electron Ion Collider}
  \vspace{1.2cm}

Rabah Abdul Khalek$^{1,2}$,
Jacob J. Ethier$^{1,2}$,
Emanuele R. Nocera$^{2,3}$, 
and Juan Rojo$^{1,2}$
\vspace{1.0cm}
 
{\it \small 
~$^1$ Department of Physics and Astronomy, VU Amsterdam, 
 1081 HV Amsterdam,\\[0.1cm]
~$^2$ Nikhef Theory Group, Science Park 105, 
 1098 XG Amsterdam, The Netherlands\\[0.1cm]
 ~$^3$ The Higgs Centre for Theoretical Physics,\\
 University of Edinburgh, JCMB, KB, Mayfield Rd, Edinburgh EH9 3FD,
 United Kingdom
}
\end{center}

\vspace{0.7cm}

\begin{abstract}
  \noindent We quantify the impact of unpolarized lepton-proton and
  lepton-nucleus inclusive deep-inelastic scattering (DIS) cross section
  measurements from the future Electron-Ion Collider (EIC) on the proton
  and nuclear parton distribution functions (PDFs). To this purpose we include
  neutral- and charged-current DIS pseudodata in a self-consistent set of
  proton and nuclear global PDF determinations based on the NNPDF methodology.
  We demonstrate that the EIC measurements will reduce the uncertainty of the
  light quark PDFs of the proton at large values of the momentum fraction $x$,
  and, more significantly, of the quark and gluon PDFs of heavy nuclei,
  especially at small and large $x$. We illustrate
  the implications of the improved precision
  of nuclear PDFs for the interaction of ultra-high
  energy cosmic neutrinos with matter.
\end{abstract}

\myparagraph{Introduction} The construction of an Electron-Ion Collider
(EIC)~\cite{Accardi:2012qut,Aschenauer:2017jsk} has been recently approved by
the United States Department of Energy at Brookhaven National Laboratory, and
could record the first scattering events as early as 2030. By colliding
(polarized) electron or positron beams with proton or ion beams for a range of
center-of-mass energies, the EIC will perform key measurements to investigate
quantum chromodynamics (QCD) at the intensity frontier. These measurements will
be fundamental to understand how partons are distributed in position and
momentum spaces within a proton, how the proton spin originates from the spin
and the dynamics of partons, how the nuclear medium modifies partonic
interactions, and whether gluons saturate within heavy nuclei.

In this paper we focus on one important class of EIC measurements, namely
inclusive cross sections for unpolarized lepton-proton and lepton-nucleus
deep-inelastic scattering (DIS). In particular we study how such data could
improve the determination of the unpolarized proton and nuclear parton
distribution functions (PDFs)~\cite{Ethier:2020way} by incorporating suitable
pseudodata in a self-consistent set of PDF fits based on the NNPDF methodology
(see Ref.~\cite{Ball:2014uwa} and references therein for a comprehensive
description). The unique ability of an EIC to measure inclusive DIS cross
sections consistently for the proton and a wide range of nuclei will be
exploited also to update the proton PDFs used as a boundary condition in the
nuclear PDF fit. This feature distinguishes our analysis from previous
studies~\cite{Aschenauer:2017oxs,AbdulKhalek:2019mzd}, and may be extended to a
simultaneous determination of proton and nuclear PDFs in the future. The
results presented in this work integrate those contained in Sects.~7.1.1 and
7.3.3 of the upcoming EIC Yellow Report~\cite{AbdulKhalek:2021gbh}. They systematically
account for the impact of projected inclusive DIS measurements at an EIC on the
unpolarized proton PDFs for the first time (for projected semi-inclusive DIS
measurements see Ref.~\cite{Aschenauer:2019kzf}), and supersede a previous
NNPDF analysis of the impact of EIC measurements on nuclear
PDFs~\cite{AbdulKhalek:2019mzd}. Similar studies for polarized PDFs have been
performed elsewhere~\cite{Aschenauer:2012ve,Aschenauer:2013iia,
  Aschenauer:2015ata,Aschenauer:2020pdk}, including in the NNPDF
framework~\cite{Ball:2013tyh}.

The structure of this paper is as follows. We first describe how EIC pseudodata
are generated. We then study how they affect the proton and nuclear PDFs once
they are fitted. Lastly, we illustrate how an updated determination of nuclear
PDFs can affect QCD at the cosmic frontier, in particular predictions for
the interactions of highly-energetic neutrinos with matter as they propagate
through Earth towards large-volume detectors.

\myparagraph{Pseudodata generation}
In this analysis we use the same pseudodata as in the EIC Yellow
Report~\cite{AbdulKhalek:2021gbh}, see in particular Sect.~8.1. In the case of lepton-proton
DIS, they consist of several sets of data points corresponding to either the
neutral-current (NC) or the charged-current (CC) DIS reduced cross sections,
$\sigma^{\rm NC}$ and $\sigma^{\rm CC}$, respectively. See, {\it e.g.}, Eqs.~(7)
and (10) in Ref.~\cite{Ball:2008by} for their definition.
Both electron and positron beams are considered, for various forecast energies
of the lepton and proton beams. In the case of lepton-nucleus DIS, the
pseudodata correspond only to NC DIS cross sections, see, {\it e.g.}, the
discussion in Sect.~2.1 of Ref.~\cite{AbdulKhalek:2019mzd} for their definition.
Both electron and positron beams are considered in conjunction with a deuteron
beam; only an electron beam is instead considered for other ions, namely
$^4$He, $^{12}$C, $^{40}$Ca, $^{64}$Cu, and $^{197}$Au.
A momentum transfer $Q^2>1$~GeV$^2$, a squared invariant mass of the system
$W^2>10$~GeV$^2$ and a fractional energy of the virtual particle exchanged in
the process $0.01\leq y \leq 0.95$ are assumed in all of the above cases,
consistently with the detector requirements outlined in Sect.8.1
of~\cite{AbdulKhalek:2021gbh}.

The pseudodata distribution is assumed to be multi-Gaussian, as in the case
of real data. It is therefore uniquely identified by a vector of mean values
$\bm{\mu}$ and a covariance matrix $\bm{\Sigma}$, for which the following
assumptions are made. The mean values correspond to the theoretical expectations
$\bm{t}$ of the DIS cross sections obtained with a {\it true} underlying set of
PDFs, and smeared by normal random numbers $\bm{r}$ sampled from the covariance
matrix such that $\bm{\mu = t + r \Sigma}$. Specifically we use a recent
variant~\cite{Faura:2020oom} of the NNPDF3.1 determination~\cite{Ball:2017nwa},
and the nNNPDF2.0 determination~\cite{AbdulKhalek:2020yuc}, for proton and
nuclear PDFs, respectively. The covariance matrix is made up of three
components, which correspond to a statistical uncertainty, an additive
uncorrelated systematic uncertainty, and a multiplicative correlated
systematic uncertainty. The statistical uncertainty is determined by assuming
an integrated luminosity $\mathcal{L}$ of 100 fb$^{-1}$ for electron-proton
NC and CC DIS, and of 10 fb$^{-1}$ in all other cases. The systematic
uncertainties are instead determined with the {\sc djangoh} event
generator~\cite{Charchula:1994kf}, which contains the Monte Carlo program
{\sc heracles}~\cite{Kwiatkowski:1990es} interfaced to
{\sc lepto}~\cite{Ingelman:1996mq}. These pieces of software collectively allow
for an account of one-loop electroweak radiative corrections and radiative
scattering. The Lund string fragmentation model, as implemented in
{\sc pythia/jetset} (see, {\it e.g.}, Ref.~\cite{Sjostrand:2019zhc} and
references therein) is used to obtain the complete hadronic final state.
The non-perturbative proton and nuclear PDF input is made available to
{\sc djangoh} by means of numeric tables corresponding to the relevant NC and
CC DIS structure functions, which were generated with
{\sc apfel}~\cite{Bertone:2013vaa} in the format of
{\sc lhapdf}~\cite{Buckley:2014ana} grids. The optimal binning of the pseudodata
is determined accordingly.

The complete set of pseudodata considered in this work is summarized in
Table~\ref{tab:pseudodata}. For each pseudodata set, we indicate the
corresponding DIS process, the number of data points $n_{\rm dat}$ before (after)
applying kinematic cuts (see below), the energy of the lepton and of the proton
or ion beams $E_\ell$ and $E_p$, the center-of-mass energy $\sqrt{s}$, the
luminosity $\mathcal{L}$, and the relative uncorrelated and correlated
systematic uncertainties (in percentage) $\sigma_u$ and $\sigma_c$.
Two different scenarios, called {\it optimistic} and {\it pessimistic}
henceforth, are considered, which differ for the number of data points and
for the size of the projected systematic uncertainties. In the case of NC
cross sections, the uncorrelated uncertainty was estimated to be 1.5\% (2.3\%)
in the optimistic (pessimistic) scenario. These uncertainties originated from a 1\%
uncertainty on the radiative corrections, and a 1\% (2\%) uncertainty due to
detector effects. The normalization uncertainty was set to 2.5\% (4.3\%). This
included a 1\% uncertainty on the integrated luminosity and a 2\% (4\%)
uncertainty due to detector effects. In the case of CC cross sections,
an uncorrelated uncertainty of 2\% was used in both the optimistic and
pessimistic scenarios, while a normalization uncertainty of 2.3\% (5.8\%)
was used in the optimistic (pessimistic) scenario. This uncertainty includes
contributions from luminosity, radiative corrections and simulation errors.

Estimating systematic uncertainties for an accelerator and a detector which
have not yet been constructed is particularly challenging. The percentages
given in Table~\ref{tab:pseudodata} build upon the experience of previous
experiments (primarily those at HERA) as well as simulation studies performed
using the EIC Handbook detector and the current EIC detector
matrix~\cite{EIC:handbook}. Relative systematic uncertainties are
estimated to be independent from the values of $x$ and $Q^2$, in contrast to
statistical uncertainties. For NC pseudodata (with $\mathcal{L}$=100~fb$^{-1}$),
systematic uncertainties are significantly larger than statistical
uncertainties in much of the probed kinematic phase space, see {\it e.g.}
Figs.~7.1 and 7.67 in~\cite{AbdulKhalek:2021gbh}. Conversely, for CC pseudodata,
systematic uncertainties are comparable to statistical uncertainties for most
of the measured kinematic space.

\begin{table}[!t]
  \footnotesize
  \centering
  \renewcommand{\arraystretch}{1.3}
  \begin{tabularx}{\linewidth}{XXXccrrcc}
    \toprule
    & \multicolumn{2}{l}{\ DIS process}
    & $n_{\rm dat}$
    & $E_\ell\times E_p$~[GeV]
    & $\sqrt{s}$~[GeV]
    & $\mathcal{L}$~[fb$^{-1}$]
    & $\sigma_u$~[\%]
    & $\sigma_c$~[\%] \\
    \midrule
      1 
    & $e^-p$
    & CC
    & 89(89)/89(89)
    & 18$\times$275
    & 140.7
    & 100
    & 2.0/2.0
    & 2.3/5.8 \\
      2
    & $e^+p$
    & CC
    & 89(89)/89(89)
    & 18$\times$275
    & 140.7
    & 10
    & 2.0/2.0
    & 2.3/5.8 \\
    \midrule
      3
    & $e^-p$
    & NC
    & 181(140)/131(107)
    & 18$\times$275
    & 140.7
    & 100
    & 1.5/2.3
    & 2.5/4.3 \\
      4 
    & 
    & 
    & 126(81)/91(70)
    & 10$\times$100
    & 63.2
    & 100
    & 1.5/2.3
    & 2.5/4.3 \\
      5
    & 
    & 
    & 116(68)/92(66)
    & 5$\times$100
    & 44.7
    & 100
    & 1.5/2.3
    & 2.5/4.3 \\
      6
    & 
    & 
    & 87(45)/76(45)
    & 5$\times$41
    & 28.6
    & 100
    & 1.5/2.3
    & 2.5/4.3 \\
      7
    & $e^+p$
    & NC
    & 181(140)/131(107)
    & 18$\times$275
    & 140.7
    & 10
    & 1.5/2.3
    & 2.5/4.3 \\
      8 
    & 
    & 
    & 126(81)/91(70)
    & 10$\times$100
    & 63.2
    & 10
    & 1.5/2.3
    & 2.5/4.3 \\
      9
    & 
    & 
    & 116(68)/92(66)
    & 5$\times$100
    & 44.7
    & 10
    & 1.5/2.3
    & 2.5/4.3 \\
      10  
    & 
    & 
    & 87(45)/76(45)
    & 5$\times$41
    & 28.6
    & 10
    & 1.5/2.3
    & 2.5/4.3 \\
    \midrule
      11
    & $e^-d$
    & NC
    & 116(92)/116(92)
    & 18$\times$110
    & 89.0
    & 10
    & 1.5/2.3
    & 2.5/4.3 \\
      12 
    &
    &
    & 107(83)/107(83)
    & 10$\times$110
    & 66.3
    & 10
    & 1.5/2.3
    & 2.5/4.3 \\
      13  
    &
    &
    & 76(45)/76(45)
    & 5$\times$41
    & 28.6
    & 10
    & 1.5/2.3
    & 2.5/4.3 \\
      14
    & $e^+d$
    & NC
    & 116(92)/116(92)
    & 18$\times$110
    & 89.0
    & 10
    & 1.5/2.3
    & 2.5/4.3 \\
      15
    &
    &
    & 107(83)/107(83)
    & 10$\times$110
    & 66.3
    & 10
    & 1.5/2.3
    & 2.5/4.3 \\
      16 
    &
    &
    & 76(45)/76(45)
    & 5$\times$41
    & 28.6
    & 10
    & 1.5/2.3
    & 2.5/4.3 \\
    \midrule
      17  
    & $e^-{^4}$He
    & NC
    & 116(92)/116(92)
    & 18$\times$110
    & 89.0
    & 10
    & 1.5/2.3
    & 2.5/4.3 \\
      18  
    &
    &
    & 107(83)/107(83)
    & 10$\times$110
    & 66.3
    & 10
    & 1.5/2.3
    & 2.5/4.3 \\
      19  
    &
    &
    & 76(45)/76(45)
    & 5$\times$41
    & 28.6
    & 10
    & 1.5/2.3
    & 2.5/4.3 \\
      20  
    & $e^-{^{12}}$C
    & NC
    & 116(92)/116(92)
    & 18$\times$110
    & 89.0
    & 10
    & 1.5/2.3
    & 2.5/4.3 \\
      21  
    &
    &
    & 107(83)/107(83)
    & 10$\times$110
    & 66.3
    & 10
    & 1.5/2.3
    & 2.5/4.3 \\
      22  
    &
    &
    & 76(45)/76(45)
    & 5$\times$41
    & 28.6
    & 10
    & 1.5/2.3
    & 2.5/4.3 \\
      23  
    & $e^-{^{40}}$Ca
    & NC
    & 116(92)/116(92)
    & 18$\times$110
    & 89.0
    & 10
    & 1.5/2.3
    & 2.5/4.3 \\
      24  
    &
    &
    & 107(83)/107(83)
    & 10$\times$110
    & 66.3
    & 10
    & 1.5/2.3
    & 2.5/4.3 \\
      25  
    &
    &
    & 76(45)/76(45)
    & 5$\times$41
    & 28.6
    & 10
    & 1.5/2.3
    & 2.5/4.3 \\
      26  
    & $e^-{^{64}}$Cu
    & NC
    & 116(92)/116(92)
    & 18$\times$110
    & 89.0
    & 10
    & 1.5/2.3
    & 2.5/4.3 \\
      27  
    &
    &
    & 107(83)/107(83)
    & 10$\times$110
    & 66.3
    & 10
    & 1.5/2.3
    & 2.5/4.3 \\
      28  
    &
    &
    & 76(45)/76(45)
    & 5$\times$41
    & 28.6
    & 10
    & 1.5/2.3
    & 2.5/4.3 \\
      29  
    & $e^-{^{197}}$Au
    & NC
    & 116(92)/116(92)
    & 18$\times$110
    & 89.0
    & 10
    & 1.5/2.3
    & 2.5/4.3 \\
      30  
    &
    &
    & 107(83)/107(83)
    & 10$\times$110
    & 66.3
    & 10
    & 1.5/2.3
    & 2.5/4.3 \\
      31  
    &
    &
    & 76(45)/76(45)
    & 5$\times$41
    & 28.6
    & 10
    & 1.5/2.3
    & 2.5/4.3 \\    
    \bottomrule
  \end{tabularx}

  \vspace{0.3cm}
  \caption{The EIC pseudodata sets considered in this work. For each of them we
    indicate the corresponding DIS process, the number of data points
    $n_{\rm dat}$ in the optimistic/pessimistic scenarios before (after)
    kinematic cuts, the energy of the lepton and of the proton or ion beams
    $E_\ell$ and $E_p$, the center-of-mass energy $\sqrt{s}$, the integrated
    luminosity $\mathcal{L}$, and the relative uncorrelated and correlated
    systematic uncertainties (in percentage) $\sigma_u$ and $\sigma_c$
    in the optimistic/pessimistic scenarios.}
  \label{tab:pseudodata}
\end{table}

\begin{figure}[!t]
 \centering
 \includegraphics[width=\textwidth,clip=true,trim=0 0cm 0 0cm]{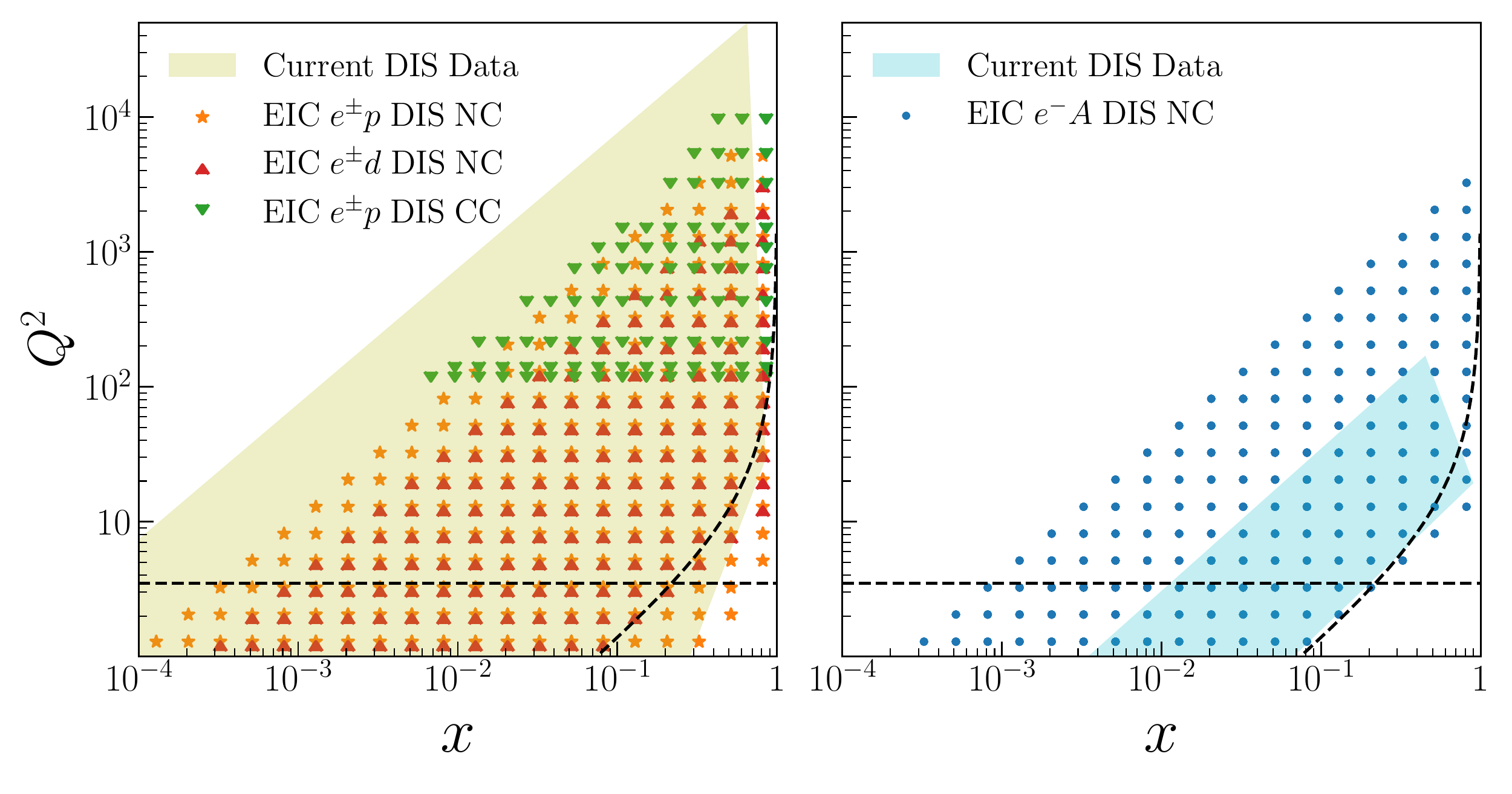}
 \caption{The expected kinematic coverage in the $(x,Q^2)$ plane of the EIC
   pseudodata for lepton-proton or lepton-deuteron (left) and lepton-nucleus
   (right panel) collisions, see Table~\ref{tab:pseudodata}. Shaded areas
   indicate the approximate kinematic coverage of the available inclusive DIS
   measurements. The dashed lines denote the kinematic cuts used in the PDF
   fits, $Q^2 \ge 3.5$ GeV$^2$ and $W^2 \ge 12.5$ GeV$^2$.}
 \label{fig:kinematics}
\end{figure}

The kinematic coverage of the EIC pseudodata in the ($x,Q^2$) plane is displayed
in Fig.~\ref{fig:kinematics} for the optimistic scenario. Pseudodata for
lepton-proton and lepton-deuteron are separated from pseudodata for electron-ion
collisions via different panels. The approximate coverage of currently available
inclusive DIS measurements is shown as a shaded area. Dashed lines correspond
to the kinematic cuts used in the PDF fits described below. From
Fig.~\ref{fig:kinematics}, we already can appreciate the relevance of the EIC
for the determination of nuclear PDFs. In this case, the EIC measurements extend
the kinematic reach of DIS by more than one order of magnitude in both $x$ and
$Q^2$. In the case of proton PDFs, instead, the EIC measurements mostly
overlap with those already available, in particular from HERA, except for a
slightly larger extension at very high $x$ and $Q^2$.

\myparagraph{Fitting procedure}
We include the pseudodata in the series of fits summarized in
Table~\ref{tab:fits}. All these fits use the NNPDF methodology. Because nuclear
PDFs are correlated with proton PDFs (the former should reduce to the latter in
the limit $A\to 1$, where $A$ is the nucleon number), and because the EIC
measurements of Table~\ref{tab:pseudodata} will affect both, we determine them
sequentially.
\begin{table}[!t]
  \footnotesize
  \centering
  \renewcommand{\arraystretch}{1.3}
\begin{tabularx}{\linewidth}{lX}
  \toprule
  Fit ID & Description \\
  \midrule
    NNPDF3.1+EIC (optimistic)
  & Same as the {\tt base} fit of~\cite{Faura:2020oom} augmented with
    the $e^\pm p$ (CC and NC) and $e^\pm d$ (NC) EIC pseudodata sets
    for the optimistic scenario. \\
    NNPDF3.1+EIC (pessimistic)
  & Same as NNPDF3.1+EIC (optimistic), but with EIC pseudodata sets for the
    pessimistic scenario. \\
  \midrule
    NNPDF3.1\_pch+EIC (optimistic)
  & Same as the proton baseline fit of~\cite{AbdulKhalek:2020yuc} augmented with
    the $e^\pm p$ (CC and NC) pseudodata sets for the optimistic scenario. \\
    NNPDF3.1\_pch+EIC (pessimistic)
  & Same as NNPDF3.1\_pch+EIC (optimistic), but with EIC pseudodata sets for
    the pessimistic scenario. \\
    nNNPDF2.0+EIC (optimistic)
  & Same as the nuclear fit of~\cite{AbdulKhalek:2020yuc} augmented with the $e^-A$ (NC) pseudodata sets (with $A=^2$d,$^4$He,$^{12}$C,
    $^{40}$Ca, $^{64}$Cu and $^{197}$Au for the optimistic scenario. \\
    nNNPDF2.0+EIC (pessimistic)
  & Same as nNNPDF2.0+EIC (optimistic), but with EIC pseudodata sets for
    the pessimistic scenario. \\
  \bottomrule
\end{tabularx}

  \caption{A summary of the fits performed in this study, see text for details.}
  \label{tab:fits}
\end{table}

First, we focus on the proton PDFs, and perform the NNPDF3.1+EIC optimistic and
pessimistic fits. These are a rerun of the {\tt base} fit of
Ref.~\cite{Faura:2020oom}, which is now augmented with the $e^\pm p$ (CC and NC)
and $e^\pm d$ (NC) EIC pseudodata sets for the optimistic and pessimistic
scenarios. As in Ref.~\cite{Ball:2017nwa,Faura:2020oom}, they are all
made of $N_{\rm rep}=100$ Monte Carlo replicas. After kinematic cuts, the fits
include a total of 5264 (5172) data points in the optimistic (pessimistic)
scenario, out of which 1286 (1194) are EIC pseudodata and 3978 are real data
(see Ref.~\cite{Faura:2020oom} for details). Kinematic cuts are the same as
in Ref.~\cite{Ball:2017nwa,Faura:2020oom}, specifically
$Q^2>3.5$~GeV$^2$ and $W^2>12.5$~GeV$^2$. These cuts, which serve the purpose
of removing a kinematic region in which potentially large higher-twist and
nuclear effects may spoil the accuracy of the PDF analysis, are more
restrictive than those used to generate the pseudodata. This fact is however
not contradictory, and reproduces what customarily happens with real data, when
different kinematic cuts are used in the experimental analysis and in a fit.
These fits are accurate to next-to-next-to-leading order (NNLO) in perturbative
QCD, they utilize the FONLL scheme~\cite{Forte:2010ta,Ball:2015tna,Ball:2015dpa}
to treat heavy quarks, and they include a parametrization of the charm PDF on
the same footing as the lighter quark PDFs. In comparison to the original
NNPDF3.1 fits~\cite{Ball:2017nwa}, a bug affecting the computation of
theoretical predictions for charged-current DIS cross sections has been
corrected, the positivity of the $F_2^c$ structure function has been enforced,
and NNLO massive corrections~\cite{Berger:2016inr,Gao:2017kkx} have been
included in the computation of neutrino-DIS structure functions.

We then focus on nuclear PDFs, and perform the NNPDF3.1\_pch+EIC and
nNNPDF2.0+EIC optimistic and pessimistic fits. These are a rerun of the
proton and nuclear baseline determinations of Ref.~\cite{AbdulKhalek:2020yuc},
augmented respectively with the $e^\pm p$ (CC and NC) and the $e^-A$ (NC),
$A=d$, $^4$He, $^{12}$C, $^{40}$Ca, $^{64}$Cu, and $^{197}$Au,
pseudodata sets for the optimistic and pessimistic scenarios.
As in Ref.~\cite{AbdulKhalek:2020yuc}, the proton (nuclear) fits are made of
$N_{\rm rep}=100$ ($N_{\rm rep}=250$) Monte Carlo replicas. After kinematic 
cuts, the NNPDF3.1\_pch+EIC fits include a total of 4147 (4055) data points
in the optimistic (pessimistic) scenario, out of which 846 (754) are EIC
pseudodata and 3301 are real data (see Ref.~\cite{AbdulKhalek:2020yuc} for
details). The nuclear fits include a total of 3007 data points, out of which
1540 are EIC pseudodata and 1467 are real data. Kinematic cuts are the same as
above, and are in turn equivalent to these used in
Refs.~\cite{Ball:2017nwa,AbdulKhalek:2020yuc}. These fits are accurate to
next-to-leading order (NLO) in perturbative QCD, and assume that charm is
generated perturbatively, consistent with Ref.~\cite{AbdulKhalek:2020yuc}.

Although the proton and nuclear PDF fits are performed independently,
they remain as consistent as possible. Most importantly, the unique
feature of an EIC to measure DIS cross sections with a comparable accuracy and
precision for a wide range of nuclei and for the proton is
key to inform the fit of nuclear PDFs as much as possible. Not only
do the measurements on nuclear targets enter the fit directly, but also the
measurements on a proton target are first used to update the necessary
baseline proton PDF determination. This feature distinguishes our work from
previous similar studies~\cite{Aschenauer:2017oxs,AbdulKhalek:2019mzd}, where
only the effect of measurements on nuclear targets were taken into account in
the determination of nuclear PDFs. A simultaneous determination of proton and
nuclear PDFs might eventually become advisable at an EIC, should the
measurements be sufficiently precise to make an independent determination
less reliable.

We also note that the pseudodata sets for a deuteron target are alternatively
included in the fit of proton PDFs or in the fit of nuclear PDFs. To avoid
double counting, they are not included in the fit of proton PDFs used
as baseline for the fit of nuclear PDFs. This choice follows the common
practice to include fixed-target DIS data on deuteron targets in fits of proton
PDFs, as done, {\it e.g.}, in NNPDF3.1 and in the variant fit used here to
generate the pseudodata. The reason being that they are essential to achieve a
good quark flavour separation. The EIC pseudodata sets for a deuteron target
are then treated, in the proton PDF fits performed here, similarly to the
fixed-target DIS data already included in NNPDF3.1. Specifically we assume
that nuclear corrections are negligible, and therefore we do not include them.
This assumption could be overcome by means of a simultaneous fit of proton and
nuclear PDFs, or by means of the iterative procedure proposed
in Ref.~\cite{Ball:2020xqw}, whereby proton and deuteron PDFs are determined by
subsequently including the uncertainties of each in the other. Any of these
approaches goes beyond the scope of this work, as they will have little
applicability in the context of pseudodata.

\myparagraph{Results} We now turn to discuss the results of the fits collected
in Table~\ref{tab:pseudodata}. As expected, the goodness of each fit 
measured by the $\chi^2$ per number of data points is comparable to that of the
fits used to generate the pseudodata. The description of each data set remains
unaltered within statistical fluctuations, and the $\chi^2$ per
number of data points for each of the new EIC pseudodata sets is of order one,
as it should by construction. In the following we therefore exclusively discuss
how the EIC pseudodata affect PDF uncertainties.

\begin{figure}[!t]
  \centering
  \includegraphics[width=\textwidth,clip=true,trim=1.5cm 0cm 1.5cm 2cm]{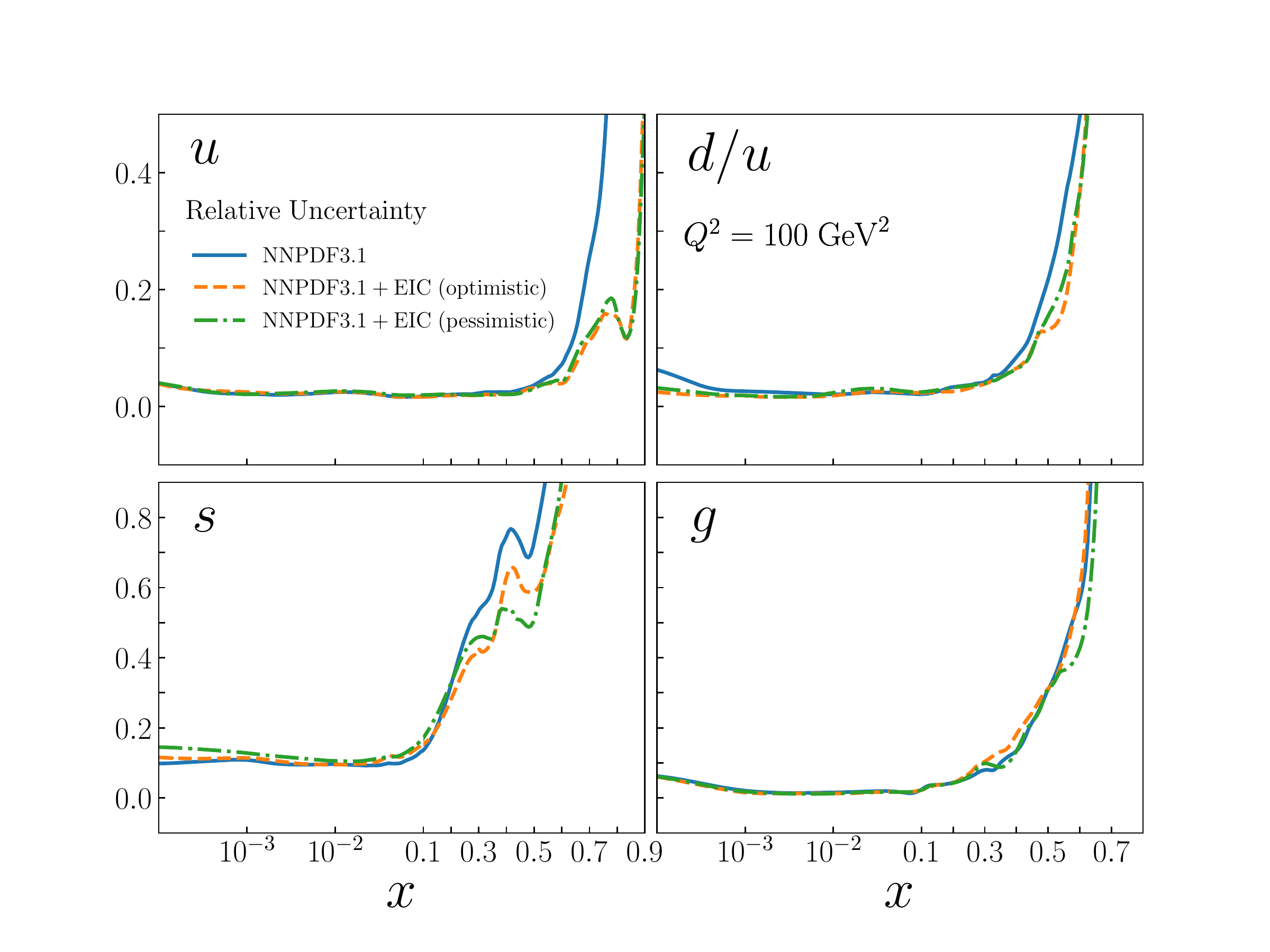}
  \caption{The relative uncertainty of the proton PDFs determined in the
  NNPDF3.1 fit variant used to generate the pseudodata, and in the
  NNPDF3.1+EIC fits, in the optimistic and pessimistic scenarios. Uncertainties
  correspond to one standard deviation and are computed as a function of $x$
  at $Q^2=100$~GeV$^2$. Only the subset of flavors (or flavor combinations) that
  are the most affected by the EIC pseudodata are shown, namely $u$,
  $d/u$, $s$ and $g$. Note the use of a log/linear scale on the $x$ axis.}
  \label{fig:ppdfs}
\end{figure}

In Fig.~\ref{fig:ppdfs} we show the relative uncertainty of the proton PDFs
in the NNPDF3.1 fit variant used to generate the pseudodata, and in the
NNPDF3.1+EIC fits, both for the optimistic and pessimistic scenarios. In each
case, uncertainties correspond to one standard deviation, and are computed
as a function of $x$ at $Q^2=100$~GeV$^2$. Only the subset of flavors (or
flavor combinations) that are the most affected by the EIC
pseudodata are shown: $u$, $d/u$, $s$ and $g$.

Fig.~\ref{fig:ppdfs} allows us to make two conclusions. First, the impact of the
EIC pseudodata is localized in the large-$x$ region, as expected from their
kinematic reach (see Fig.~\ref{fig:kinematics}). This impact is significant in
the case of the $u$ PDF, for which PDF uncertainties could be reduced by up to
a factor of two for $x\gtrsim 0.7$. The impact is otherwise moderate for the
$d/u$ PDF ratio (for which it amounts to an uncertainty reduction of about
one third for $0.5\lesssim x \lesssim 0.6$) and for the $s$ PDF (for which 
it amounts to an uncertainty reduction of about one fourth for
$0.3\lesssim x \lesssim 0.6$). The relative uncertainty of the gluon PDF, and
of other PDFs not shown in Fig.~\ref{fig:ppdfs}, remains unaffected.
These features rely on the unique ability of the EIC to perform precise DIS
measurements at large $x$ and large $Q^2$: their theoretical interpretation
remains particularly clean, as any non-perturbative large-$x$ contamination
due, {\it e.g.}, to higher-twist effects, is suppressed. This possibility
distinguishes the EIC from HERA, which had a similar reach at high $Q^2$ but a
more limited access at large-$x$, and from fixed-target experiments
(including the recent JLab-12 upgrade~\cite{Dudek:2012vr}), which can access the
high-$x$ region only at small $Q^2$. Secondly, the impact of the EIC pseudodata
does not seem to depend on the scenario considered: the reduction of PDF
uncertainties remains comparable irrespective of whether optimistic or
pessimistic pseudodata projections are included in the fits. Because the two
scenarios only differ in systematic uncertainties, we conclude that it may be
sufficient to control these to the level of precision forecast
in the pessimistic scenario.

\begin{figure}[!t]
  \centering
  \includegraphics[width=\textwidth,clip=true,trim=1.5cm 2.5cm 1.5cm 3.5cm]{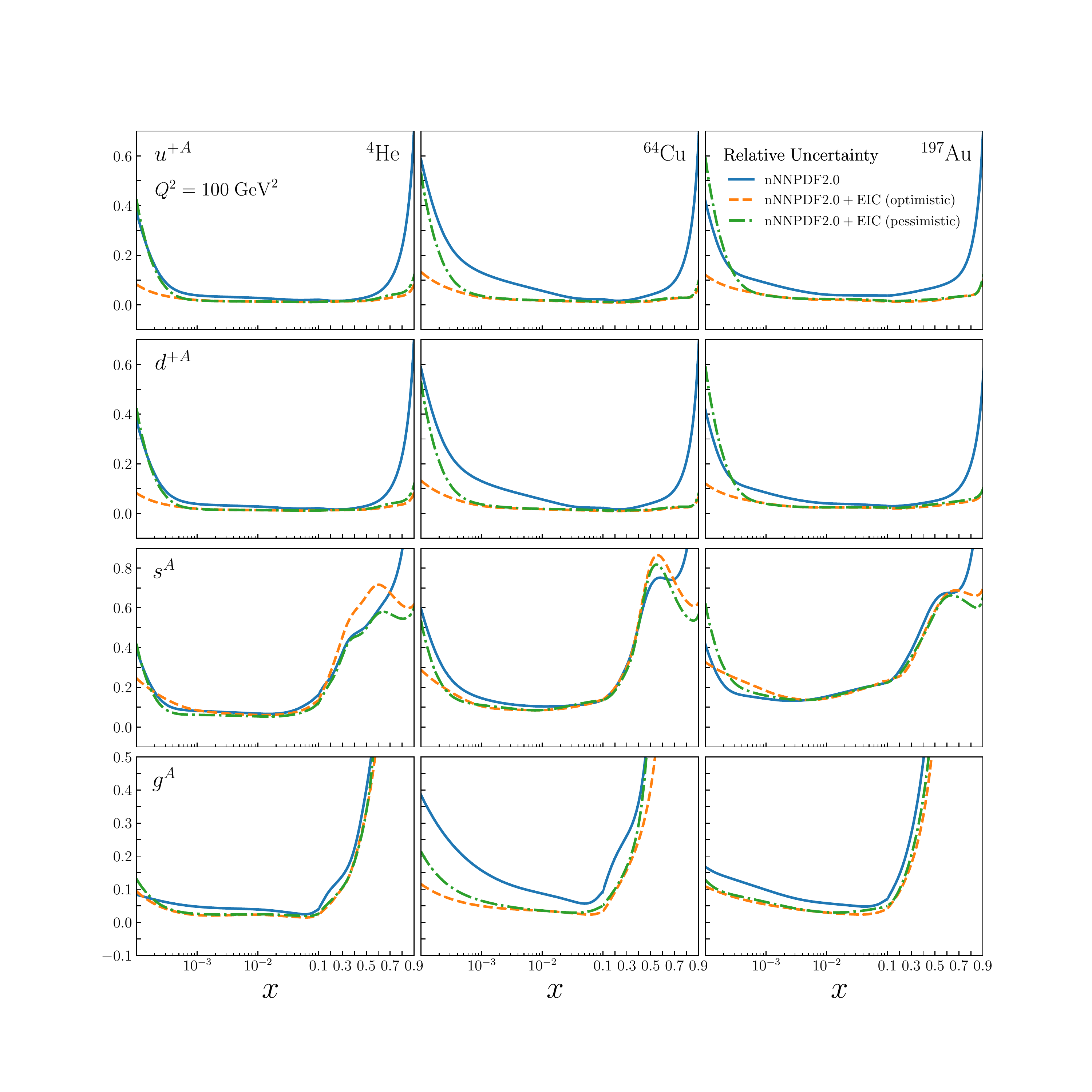}
  \caption{The relative uncertainty of the nuclear PDFs determined in the
  nNNPDF2.0 fit used to generate the pseudodata, and in the nNNPDF2.0+EIC fits,
  in the optimistic and pessimistic scenarios. Uncertainties correspond to one
  standard deviation, and are computed as a function of $x$ at
  $Q^2$=100~GeV$^2$. Results are displayed for the ions with the lowest and
  highest atomic mass, $^4$He (left) and $^{197}$Au (right), and for an
  intermediate atomic mass ion, $^{64}$Cu (middle column), and only for the
  PDF flavors that are the most affected by the EIC pseudodata: $u$, $\bar{d}$,
  $s$ and $g$. Note the use of a log/linear scale on the $x$ axis.}
  \label{fig:npdfs}
\end{figure}

A similar behavior is observed for the NNPDF3.1\_pch fits, which are therefore
not displayed. In Figs.~\ref{fig:npdfs} we show the relative uncertainty of the
nuclear PDFs in the nNNPDF2.0 fit used to generate the pseudodata, and in the
nNNPDF2.0+EIC fits, both in the pessimistic and in the optimistic scenarios.
Uncertainties correspond to one standard deviation, and
are computed as a function of $x$ at $Q^2$=100~GeV$^2$. Results
are displayed for the ions with the lowest and highest atomic mass, $^4$He and
$^{197}$Au, and for an intermediate atomic mass ion, $^{64}$Cu,
and only for the PDF flavors that are the most affected by
the EIC pseudodata: $u$, $\bar{d}$, $s$ and $g$.

From Fig.~\ref{fig:npdfs} we observe a reduction of nuclear
PDF uncertainties, due to EIC pseudodata, that varies with the nucleus, the
$x$ region considered, and the PDF. Overall, the heavier the nucleus,
the largest the reduction of PDF uncertainties. This is a consequence of the
fact that nuclear PDFs are customarily parametrized as continuous functions
of the nucleon number $A$: nuclear PDFs for $^4$He, which differ from the proton
PDF boundary by a small correction, are better constrained than nuclear PDFs
for $^{197}$Au because proton data are more abundant than data for nuclei.
In this respect, the EIC will allow one to perform a comparatively accurate
scan of the kinematic space for each nucleus individually, and, as shown in
Fig.~\ref{fig:npdfs}, to determine the PDFs of all ions with a similar
precision. The reduction of PDF uncertainties is localized in the small-$x$
region, where little or no data are currently available
(see Fig.~\ref{fig:kinematics}), and
in the large-$x$ region, where nuclear PDF benefit from the increased precision
of the baseline proton PDFs. In the case of the gluon PDF, the reduction of
uncertainties is seen for the whole range in $x$. This is a consequence of the
extended data coverage in $Q^2$, which allows one to constrain the gluon PDF
even further via perturbative evolution. As observed in the case of
proton PDFs, the fits obtained upon inclusion of the EIC pseudodata do not
significantly differ whether the optimistic or the pessimistic scenarios are
considered, except for very small values of $x$. In this case the optimistic
scenario leads to a more marked reduction of PDF uncertainties, especially for
the total PDF combinations $u^+$ and $d^+$. This feature is mainly driven by the
smaller systematic uncertainties that affect the NC pseudodata in the optimistic
scenario (about 3\%) with respect to the pessimistic one (about 5\%), see
Table~\ref{tab:pseudodata}. That is aligned with the fact that the statistical
uncertainties are comparable between the two scenarios. 

\myparagraph{Implications for neutrino astrophysics} The reduction of PDF
uncertainties due to EIC pseudodata, in particular for
nuclear PDFs, may have important phenomenological implications. Not only at the
intensity frontier, {\it e.g.} to characterize gluon saturation at small $x$,
but also at the energy frontier, {\it e.g.}, for searches of new physics that
require a precise knowledge of PDFs at high $x$, and at the cosmic frontier,
{\it e.g.}, in the detection of highly energetic neutrinos from astrophysical
sources. We conclude our paper by focusing on this last aspect. Specifically
it was shown in Ref.~\cite{Garcia:2020jwr} that the dominant source of
uncertainty in the theoretical predictions for the cross section of
neutrino-matter interactions is represented by nuclear effects.
The corresponding NC and CC inclusive DIS cross sections may differ
significantly depending on whether they are computed for neutrino-nucleon or
neutrino-nucleus interactions. The uncertainty is larger in the latter case,
because nuclear PDFs are not as precise as proton PDFs, and is such that it
encompasses the difference in central values. We revisit this statement in
light of the precise nNNPDF3.0+EIC fits.

\begin{figure}[!t]
  \centering
  \includegraphics[width=\textwidth]{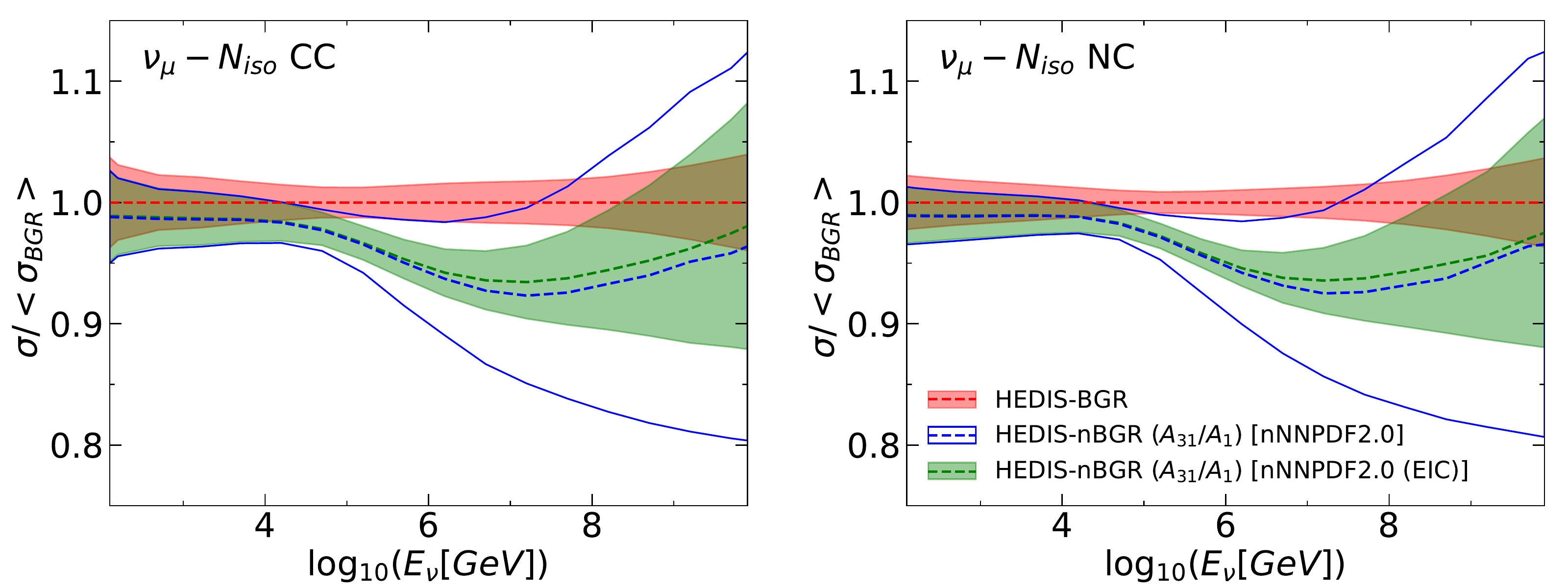}
  \caption{The CC (left) and NC (right) neutrino-nucleus DIS cross
  sections, with their one-sigma uncertainties, as a function of the neutrino
  energy $E_{\nu}$. Predictions correspond to the HEDIS-BGR
  computation~\cite{Garcia:2020jwr} with the proton PDF of~\cite{Gauld:2016kpd},
  and with the nNNPDF2.0 and nNNPDF2.0+EIC nuclear PDFs. They are all
  normalized to the central value of the proton results. See text for details.}
  \label{fig:UHExsec}
\end{figure}

\begin{figure}[!t]
  \centering
  \includegraphics[width=\textwidth]{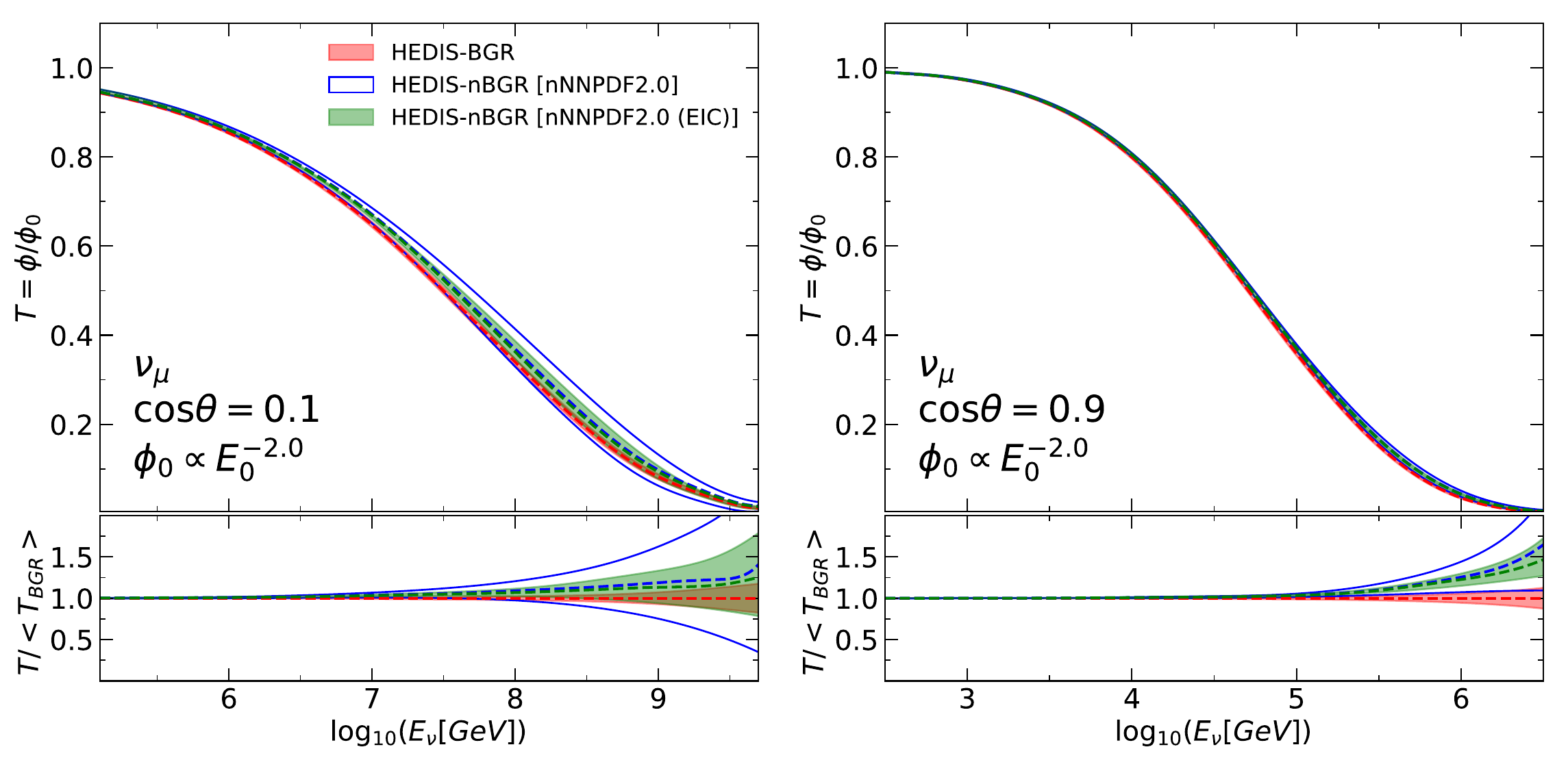}
  \caption{The transmission coefficient $T$ for muonic neutrinos as a function
  of the neutrino energy $E_\nu$ and for two values of the nadir angle $\theta$.
  Predictions correspond to the computation of~\cite{Garcia:2020jwr} with the
  proton PDF of~\cite{Gauld:2016kpd}, and with the nNNPDF2.0 and nNNPDF2.0+EIC
  nuclear PDFs. They are all normalized to the central value of the proton
  results. See text for details.}
  \label{fig:attenuation}
\end{figure}

In Fig.~\ref{fig:UHExsec} we show the CC (left) and NC (right)
neutrino-nucleus inclusive DIS cross sections, with their one-sigma PDF
uncertainties, as a function of the neutrino energy $E_\nu$. Moreover, in
Fig.~\ref{fig:attenuation} we show the transmission coefficient $T$ for muonic
neutrinos, defined as the ratio between the incoming neutrino flux $\Phi_0$ and
the flux arriving at the detector volume $\Phi$ (see Eq.~(3.1) and the ensuing
discussion in Ref.~\cite{Garcia:2020jwr} for details); $T$ is displayed for two
values of the nadir angle $\theta$ as a function of the neutrino energy $E_\nu$.
In both cases, we compare predictions obtained with the calculation presented
in Refs.~\cite{Bertone:2018dse,Garcia:2020jwr} and implemented in
{\sc hedis}~\cite{Brown:1971qr}. For a proton target the prediction is made
with the proton PDF set determined in Ref.~\cite{Gauld:2016kpd},
a variant of the NNPDF3.1 PDF set in which small-$x$ resummation
effects~\cite{Ball:2017otu} and additional constraints from $D$-meson
production measurements in proton-proton collisions at 5,7 and
13~TeV~\cite{Aaij:2013mga,Aaij:2015bpa,Aaij:2016jht} have been included.
This prediction is labeled HEDIS-BGR in
Figs.~\ref{fig:UHExsec}-\ref{fig:attenuation}. For a nuclear
target ($A=31$ is adopted as in Ref.~\cite{Garcia:2020jwr}), the prediction is
made alternatively with the nNNPDF2.0 and the nNNPDF2.0+EIC (optimistic) PDFs.
The corresponding predictions are labeled HEDIS-nBGR [nNNPDF2.0] and HEDIS-nBGR
[nNNPDF2.0 (EIC)] in Figs.~\ref{fig:UHExsec}-\ref{fig:attenuation}. Predictions
are all normalized to the central value of the proton result. In comparison to
nNNPDF2.0, the effect of the EIC pseudodata is seen to reduce the uncertainty
of the prediction for a nuclear target by roughly a factor of two for
$E_\nu\simeq 10^6$~GeV. The reduced uncertainty no longer encompasses the
difference between predictions obtained on a proton or on a nuclear target,
except in the case of an attenuation rate computed with a large nadir angle.
Furthermore, this reduction extends to much larger neutrino energy
($E_\nu \gtrsim 10^7$), beyond the EIC-sensitive $x$-region
of the PDFs. We believe this to be partly due to DGLAP evolution and sum rules
that smoothen the low-$x$ PDF behavior, but also potentially a consequence of
the factorisation approximation used to account for nuclear corrections in the
ultra high-energy cross-sections highlighted by Eqs.~(5.2, 5.3) in
Ref.~\cite{Garcia:2020jwr}.

\myparagraph{Summary} In this paper we have quantified the impact that
unpolarized lepton-proton and lepton-nucleus inclusive DIS cross section
measurements at the future EIC will have on the unpolarized proton and nuclear
PDFs. In particular, we have extended the NNPDF3.1 and nNNPDF2.0 global
analyses by including suitable NC and CC DIS pseudodata corresponding to a
variety of nuclei and center-of-mass energies. Two different scenarios,
optimistic and pessimistic, have been considered for the projected systematic
uncertainties of the pseudodata. We have found that the EIC could reduce the
uncertainty of the light quark PDFs of the proton at large $x$, and, more
significantly, the quark and gluon PDF uncertainties for nuclei in a wide range
of atomic mass $A$ values both at small and large $x$. In general the size of
this reduction turns out to be similar for both the optimistic and
pessimistic scenarios. We therefore conclude that it may be sufficient to
control experimental uncertainties to the level of precision forecast in the
latter scenario. Lastly, we have illustrated how theoretical predictions
obtained with nuclear PDFs constrained by EIC data will improve the modelling
of the interactions of ultra-high energy cosmic neutrinos with matter.
In particular we have demonstrated that nuclear PDF uncertainties may
no longer encompass the difference between predictions obtained on a proton and
on a nuclear target. This fact highlights the increasing importance of carefully
accounting for nuclear PDF effects in high-energy neutrino astrophysics.

Further phenomenological implications could be investigated
in the future, for instance whether a simultaneous determination of proton and
nuclear PDFs can improve the constraints provided by the EIC data in comparison
to the self-consistent strategy adopted in this paper, or the extent to which
semi-inclusive DIS (SIDIS) data can further improve both proton and nuclear
PDF determinations. The PDF sets discussed in this work are available in the
LHAPDF format~\cite{Buckley:2014ana} from the NNPDF website:
\begin{center}
\url{http://nnpdf.mi.infn.it/for-users/nnnpdf2-0eic/}
\end{center}

\section*{Acknowledgements}
We thank the conveners of the Inclusive Reaction Working Group for allowing us
to utilize the EIC pseudodata generated for the upcoming EIC Yellow report, and
our colleagues in the EIC User Group for many useful discussions. We thank
Alfonso Garcia for having provided us with the numerical results displayed
in Figs.~\ref{fig:UHExsec}-\ref{fig:attenuation}. The work of R.~A.~K., J.~E.
and J.~R. is partially supported by the Netherlands Organization for Scientific
Research (NWO). E.~R.~N. is supported by the UK STFC grant ST/T000600/1 and was
also supported by the European Commission through the Marie Sk\l odowska-Curie
Action ParDHonSFFs.TMDs (grant number 752748).

\providecommand{\href}[2]{#2}\begingroup\raggedright\endgroup


\begin{thebibliography}{10}

  \bibitem{Accardi:2012qut}
  A.~Accardi et~al., {\it {Electron Ion Collider: The Next QCD Frontier}:
    {Understanding the glue that binds us all}},  {\em Eur. Phys. J. A} {\bf 52}
    (2016), no.~9 268, [\href{http://arxiv.org/abs/1212.1701}{{\tt
    arXiv:1212.1701}}].
  
  \bibitem{Aschenauer:2017jsk}
  E.~Aschenauer, S.~Fazio, J.~Lee, H.~Mantysaari, B.~Page, B.~Schenke,
    T.~Ullrich, R.~Venugopalan, and P.~Zurita, {\it {The electron\textendash{}ion
    collider: assessing the energy dependence of key measurements}},  {\em Rept.
    Prog. Phys.} {\bf 82} (2019), no.~2 024301,
    [\href{http://arxiv.org/abs/1708.01527}{{\tt arXiv:1708.01527}}].
  
  \bibitem{Ethier:2020way}
  J.~J. Ethier and E.~R. Nocera, {\it {Parton Distributions in Nucleons and
    Nuclei}},  {\em Ann. Rev. Nucl. Part. Sci.} (2020), no.~70 1--34,
    [\href{http://arxiv.org/abs/2001.07722}{{\tt arXiv:2001.07722}}].
  
  \bibitem{Ball:2014uwa}
  {\bf NNPDF} Collaboration, R.~D. Ball et~al., {\it {Parton distributions for
    the LHC Run II}},  {\em JHEP} {\bf 04} (2015) 040,
    [\href{http://arxiv.org/abs/1410.8849}{{\tt arXiv:1410.8849}}].
  
  \bibitem{Aschenauer:2017oxs}
  E.~Aschenauer, S.~Fazio, M.~Lamont, H.~Paukkunen, and P.~Zurita, {\it {Nuclear
    Structure Functions at a Future Electron-Ion Collider}},  {\em Phys. Rev. D}
    {\bf 96} (2017), no.~11 114005, [\href{http://arxiv.org/abs/1708.05654}{{\tt
    arXiv:1708.05654}}].
  
  \bibitem{AbdulKhalek:2019mzd}
  {\bf NNPDF} Collaboration, R.~Abdul~Khalek, J.~J. Ethier, and J.~Rojo, {\it
    {Nuclear parton distributions from lepton-nucleus scattering and the impact
    of an electron-ion collider}},  {\em Eur. Phys. J. C} {\bf 79} (2019), no.~6
    471, [\href{http://arxiv.org/abs/1904.00018}{{\tt arXiv:1904.00018}}].
  
  \bibitem{AbdulKhalek:2021gbh}
  R.~Abdul~Khalek et~al., {\it {Science Requirements and Detector Concepts for
    the Electron-Ion Collider: EIC Yellow Report}},
    \href{http://arxiv.org/abs/2103.05419}{{\tt arXiv:2103.05419}}.
  
  \bibitem{Aschenauer:2019kzf}
  E.~C. Aschenauer, I.~Borsa, R.~Sassot, and C.~Van~Hulse, {\it {Semi-inclusive
    Deep-Inelastic Scattering, Parton Distributions and Fragmentation Functions
    at a Future Electron-Ion Collider}},  {\em Phys. Rev. D} {\bf 99} (2019),
    no.~9 094004, [\href{http://arxiv.org/abs/1902.10663}{{\tt
    arXiv:1902.10663}}].
  
  \bibitem{Aschenauer:2012ve}
  E.~C. Aschenauer, R.~Sassot, and M.~Stratmann, {\it {Helicity Parton
    Distributions at a Future Electron-Ion Collider: A Quantitative Appraisal}},
    {\em Phys. Rev. D} {\bf 86} (2012) 054020,
    [\href{http://arxiv.org/abs/1206.6014}{{\tt arXiv:1206.6014}}].
  
  \bibitem{Aschenauer:2013iia}
  E.~C. Aschenauer, T.~Burton, T.~Martini, H.~Spiesberger, and M.~Stratmann, {\it
    {Prospects for Charged Current Deep-Inelastic Scattering off Polarized
    Nucleons at a Future Electron-Ion Collider}},  {\em Phys. Rev. D} {\bf 88}
    (2013) 114025, [\href{http://arxiv.org/abs/1309.5327}{{\tt
    arXiv:1309.5327}}].
  
  \bibitem{Aschenauer:2015ata}
  E.~C. Aschenauer, R.~Sassot, and M.~Stratmann, {\it {Unveiling the Proton Spin
    Decomposition at a Future Electron-Ion Collider}},  {\em Phys. Rev. D} {\bf
    92} (2015), no.~9 094030, [\href{http://arxiv.org/abs/1509.06489}{{\tt
    arXiv:1509.06489}}].
  
  \bibitem{Aschenauer:2020pdk}
  I.~Borsa, G.~Lucero, R.~Sassot, E.~C. Aschenauer, and A.~S. Nunes, {\it
    {Revisiting helicity parton distributions at a future electron-ion
    collider}},  {\em Phys. Rev. D} {\bf 102} (2020), no.~9 094018,
    [\href{http://arxiv.org/abs/2007.08300}{{\tt arXiv:2007.08300}}].
  
  \bibitem{Ball:2013tyh}
  {\bf NNPDF} Collaboration, R.~D. Ball, S.~Forte, A.~Guffanti, E.~R. Nocera,
    G.~Ridolfi, and J.~Rojo, {\it {Polarized Parton Distributions at an
    Electron-Ion Collider}},  {\em Phys. Lett. B} {\bf 728} (2014) 524--531,
    [\href{http://arxiv.org/abs/1310.0461}{{\tt arXiv:1310.0461}}].
  
  \bibitem{Ball:2008by}
  {\bf NNPDF} Collaboration, R.~D. Ball, L.~Del~Debbio, S.~Forte, A.~Guffanti,
    J.~I. Latorre, A.~Piccione, J.~Rojo, and M.~Ubiali, {\it {A Determination of
    parton distributions with faithful uncertainty estimation}},  {\em Nucl.
    Phys. B} {\bf 809} (2009) 1--63, [\href{http://arxiv.org/abs/0808.1231}{{\tt
    arXiv:0808.1231}}]. [Erratum: Nucl.Phys.B 816, 293 (2009)].
  
  \bibitem{Faura:2020oom}
  F.~Faura, S.~Iranipour, E.~R. Nocera, J.~Rojo, and M.~Ubiali, {\it {The
    Strangest Proton?}},  {\em Eur. Phys. J. C} {\bf 80} (2020), no.~12 1168,
    [\href{http://arxiv.org/abs/2009.00014}{{\tt arXiv:2009.00014}}].
  
  \bibitem{Ball:2017nwa}
  {\bf NNPDF} Collaboration, R.~D. Ball et~al., {\it {Parton distributions from
    high-precision collider data}},  {\em Eur. Phys. J. C} {\bf 77} (2017),
    no.~10 663, [\href{http://arxiv.org/abs/1706.00428}{{\tt arXiv:1706.00428}}].
  
  \bibitem{AbdulKhalek:2020yuc}
  R.~Abdul~Khalek, J.~J. Ethier, J.~Rojo, and G.~van Weelden, {\it {nNNPDF2.0:
    Quark Flavor Separation in Nuclei from LHC Data}},  {\em JHEP} {\bf 09}
    (2020) 183, [\href{http://arxiv.org/abs/2006.14629}{{\tt arXiv:2006.14629}}].
  
  \bibitem{Charchula:1994kf}
  K.~Charchula, G.~Schuler, and H.~Spiesberger, {\it {Combined QED and QCD
    radiative effects in deep inelastic lepton - proton scattering: The Monte
    Carlo generator DJANGO6}},  {\em Comput. Phys. Commun.} {\bf 81} (1994)
    381--402.
  
  \bibitem{Kwiatkowski:1990es}
  A.~Kwiatkowski, H.~Spiesberger, and H.~Mohring, {\it {Heracles: An Event
    Generator for $e p$ Interactions at \{HERA\} Energies Including Radiative
    Processes: Version 1.0}},  {\em Comput. Phys. Commun.} {\bf 69} (1992)
    155--172.
  
  \bibitem{Ingelman:1996mq}
  G.~Ingelman, A.~Edin, and J.~Rathsman, {\it {LEPTO 6.5: A Monte Carlo generator
    for deep inelastic lepton - nucleon scattering}},  {\em Comput. Phys.
    Commun.} {\bf 101} (1997) 108--134,
    [\href{http://arxiv.org/abs/hep-ph/9605286}{{\tt hep-ph/9605286}}].
  
  \bibitem{Sjostrand:2019zhc}
  T.~Sj\"ostrand, {\it {The PYTHIA Event Generator: Past, Present and Future}},
    {\em Comput. Phys. Commun.} {\bf 246} (2020) 106910,
    [\href{http://arxiv.org/abs/1907.09874}{{\tt arXiv:1907.09874}}].
  
  \bibitem{Bertone:2013vaa}
  V.~Bertone, S.~Carrazza, and J.~Rojo, {\it {APFEL: A PDF Evolution Library with
    QED corrections}},  {\em Comput. Phys. Commun.} {\bf 185} (2014) 1647--1668,
    [\href{http://arxiv.org/abs/1310.1394}{{\tt arXiv:1310.1394}}].
  
  \bibitem{Buckley:2014ana}
  A.~Buckley, J.~Ferrando, S.~Lloyd, K.~Nordstr\"om, B.~Page, M.~R\"ufenacht,
    M.~Sch\"onherr, and G.~Watt, {\it {LHAPDF6: parton density access in the LHC
    precision era}},  {\em Eur. Phys. J. C} {\bf 75} (2015) 132,
    [\href{http://arxiv.org/abs/1412.7420}{{\tt arXiv:1412.7420}}].
  
  \bibitem{EIC:handbook}
  See
    \href{http://eicug.org/web/sites/default/files/EIC_HANDBOOK_v1.2.pdf}{http://eicug.org/web/sites/default/files/EIC\_HANDBOOK\_v1.2.pdf}.
  
  \bibitem{Forte:2010ta}
  S.~Forte, E.~Laenen, P.~Nason, and J.~Rojo, {\it {Heavy quarks in
    deep-inelastic scattering}},  {\em Nucl. Phys. B} {\bf 834} (2010) 116--162,
    [\href{http://arxiv.org/abs/1001.2312}{{\tt arXiv:1001.2312}}].
  
  \bibitem{Ball:2015tna}
  R.~D. Ball, V.~Bertone, M.~Bonvini, S.~Forte, P.~Groth~Merrild, J.~Rojo, and
    L.~Rottoli, {\it {Intrinsic charm in a matched general-mass scheme}},  {\em
    Phys. Lett. B} {\bf 754} (2016) 49--58,
    [\href{http://arxiv.org/abs/1510.00009}{{\tt arXiv:1510.00009}}].
  
  \bibitem{Ball:2015dpa}
  R.~D. Ball, M.~Bonvini, and L.~Rottoli, {\it {Charm in Deep-Inelastic
    Scattering}},  {\em JHEP} {\bf 11} (2015) 122,
    [\href{http://arxiv.org/abs/1510.02491}{{\tt arXiv:1510.02491}}].
  
  \bibitem{Berger:2016inr}
  E.~L. Berger, J.~Gao, C.~S. Li, Z.~L. Liu, and H.~X. Zhu, {\it {Charm-Quark
    Production in Deep-Inelastic Neutrino Scattering at Next-to-Next-to-Leading
    Order in QCD}},  {\em Phys. Rev. Lett.} {\bf 116} (2016), no.~21 212002,
    [\href{http://arxiv.org/abs/1601.05430}{{\tt arXiv:1601.05430}}].
  
  \bibitem{Gao:2017kkx}
  J.~Gao, {\it {Massive charged-current coefficient functions in deep-inelastic
    scattering at NNLO and impact on strange-quark distributions}},  {\em JHEP}
    {\bf 02} (2018) 026, [\href{http://arxiv.org/abs/1710.04258}{{\tt
    arXiv:1710.04258}}].
  
  \bibitem{Ball:2020xqw}
  R.~D. Ball, E.~R. Nocera, and R.~L. Pearson, {\it {Deuteron Uncertainties in
    the Determination of Proton PDFs}},
    \href{http://arxiv.org/abs/2011.00009}{{\tt arXiv:2011.00009}}.
  
  \bibitem{Dudek:2012vr}
  J.~Dudek et~al., {\it {Physics Opportunities with the 12 GeV Upgrade at
    Jefferson Lab}},  {\em Eur. Phys. J. A} {\bf 48} (2012) 187,
    [\href{http://arxiv.org/abs/1208.1244}{{\tt arXiv:1208.1244}}].
  
  \bibitem{Garcia:2020jwr}
  A.~Garcia, R.~Gauld, A.~Heijboer, and J.~Rojo, {\it {Complete predictions for
    high-energy neutrino propagation in matter}},  {\em JCAP} {\bf 09} (2020)
    025, [\href{http://arxiv.org/abs/2004.04756}{{\tt arXiv:2004.04756}}].
  
  \bibitem{Bertone:2018dse}
  V.~Bertone, R.~Gauld, and J.~Rojo, {\it {Neutrino Telescopes as QCD
    Microscopes}},  {\em JHEP} {\bf 01} (2019) 217,
    [\href{http://arxiv.org/abs/1808.02034}{{\tt arXiv:1808.02034}}].
  
  \bibitem{Brown:1971qr}
  R.~Brown, {\it {Intermediate boson. i. theoretical production cross-sections in
    high-energy neutrino and muon experiments}},  {\em Phys. Rev. D} {\bf 3}
    (1971) 207--223.
  
  \bibitem{Gauld:2016kpd}
  R.~Gauld and J.~Rojo, {\it {Precision determination of the small-$x$ gluon from
    charm production at LHCb}},  {\em Phys. Rev. Lett.} {\bf 118} (2017), no.~7
    072001, [\href{http://arxiv.org/abs/1610.09373}{{\tt arXiv:1610.09373}}].
  
  \bibitem{Ball:2017otu}
  R.~D. Ball, V.~Bertone, M.~Bonvini, S.~Marzani, J.~Rojo, and L.~Rottoli, {\it
    {Parton distributions with small-x resummation: evidence for BFKL dynamics in
    HERA data}},  {\em Eur. Phys. J. C} {\bf 78} (2018), no.~4 321,
    [\href{http://arxiv.org/abs/1710.05935}{{\tt arXiv:1710.05935}}].
  
  \bibitem{Aaij:2013mga}
  {\bf LHCb} Collaboration, R.~Aaij et~al., {\it {Prompt charm production in pp
    collisions at sqrt(s)=7 TeV}},  {\em Nucl. Phys. B} {\bf 871} (2013) 1--20,
    [\href{http://arxiv.org/abs/1302.2864}{{\tt arXiv:1302.2864}}].
  
  \bibitem{Aaij:2015bpa}
  {\bf LHCb} Collaboration, R.~Aaij et~al., {\it {Measurements of prompt charm
    production cross-sections in $pp$ collisions at $ \sqrt{s}=13 $ TeV}},  {\em
    JHEP} {\bf 03} (2016) 159, [\href{http://arxiv.org/abs/1510.01707}{{\tt
    arXiv:1510.01707}}]. [Erratum: JHEP 09, 013 (2016), Erratum: JHEP 05, 074
    (2017)].
  
  \bibitem{Aaij:2016jht}
  {\bf LHCb} Collaboration, R.~Aaij et~al., {\it {Measurements of prompt charm
    production cross-sections in pp collisions at $ \sqrt{s}=5 $ TeV}},  {\em
    JHEP} {\bf 06} (2017) 147, [\href{http://arxiv.org/abs/1610.02230}{{\tt
    arXiv:1610.02230}}].
  
  \end{thebibliography}
\end{document}